\documentclass[%
 aip,
 amsmath,amssymb,
 reprint,%
]{revtex4-1}

\usepackage{graphicx}
\usepackage{dcolumn}
\usepackage{bm}

\usepackage[utf8]{inputenc}
\usepackage[T1]{fontenc}
\usepackage{mathptmx}
\usepackage{etoolbox}
\usepackage{xcolor}
\usepackage{svg}
\usepackage{appendix}
\usepackage{amsmath}
\usepackage{subcaption}
\usepackage{soul}
\makeatletter
\def\@email#1#2{%
 \endgroup
 \patchcmd{\titleblock@produce}
  {\frontmatter@RRAPformat}
  {\frontmatter@RRAPformat{\produce@RRAP{*#1\href{mailto:#2}{#2}}}\frontmatter@RRAPformat}
  {}{}
}%
\makeatother
\begin{document}

\preprint{AIP/123-QED}

\title[Low-dimensional Dynamics of the Social Compass Model]{Low-dimensional Dynamics of the Social Compass Model}
\author{Corbit R. Sampson}
\email{Corbit.sampson@gmail.com.}
 \affiliation{Department of Applied Mathematics, University of Colorado at Boulder, Colorado 80309, USA.}
\author{Juan G. Restrepo}
\affiliation{Department of Applied Mathematics, University of Colorado at Boulder, Colorado 80309, USA.
}%

\date{\today}

\begin{abstract}
The {\it social compass model} has been recently proposed as a model for depolarization in populations where individuals have multiple, possibly correlated, opinions. Previous work has focused on the steady state of this model, but has not addressed the dynamics leading to depolarization. We show that the macroscopic dynamics of the social compass model can be described using the Ott-Antonsen Ansatz and that, for initially clustered opinions, the resulting equations reduce to a finite-dimensional system of ordinary differential equations. We study the linear stability of the polarized state and find a dispersion relation for the growth rate of perturbations from this state. We find that the critical coupling for depolarization depends only on the first inverse moment of the conviction distribution, whereas the rate of depolarization depends on higher moments. Consequently, conviction distributions with the same critical coupling can exhibit vastly different depolarization timescales. We also demonstrate how our analysis can be extended to study depolarization in the presence of community structure. 
\end{abstract}

\maketitle

\begin{quotation}
Models of opinion dynamics are often restricted to agents holding only one opinion. In practice, agents can hold multiple opinions, and these opinions can be correlated. The social compass model was recently proposed to study how polarization in populations of agents with two potentially correlated opinions disappears as the coupling between agents becomes stronger. Previous results mostly addressed static behavior. Here we develop a dynamical theory of the social compass model based on the Ott–Antonsen Ansatz, allowing us to describe the evolution of macroscopic variables and determine how rapidly depolarization occurs. We show that heterogeneity in agents' convictions can strongly affect depolarization timescales, even when populations with different conviction distributions depolarize at the same critical coupling strength. We further illustrate how this framework can be extended to study depolarization in populations with community structure.
\end{quotation}

\section{\label{sec:intro} Introduction}

The ability of a population to reach consensus is an important aspect of social organization and underlies many collective phenomena, including the establishment of social norms, political agreement, and cultural practices \cite{lewis2008convention,hechter2009theories,baronchelli2018emergence}. In practice, however, social systems rarely exist in a state of complete consensus \cite{inglehart1998human}. Instead, populations often split into groups that hold distinct opinions, a phenomenon commonly referred to as fragmentation \cite{hegselmann2019consensus}. In more extreme cases, these groups may adopt opposing and strongly held views, leading to opinion polarization.

Opinion polarization has been observed in many contexts, including political and religious ideologies \cite{perry2022american,mccoy2018polarization}, race \cite{montalvo2005ethnic}, and climate change \cite{mccright2011politicization}. Because polarization is often associated with increased social division and a reduced ability to address collective challenges, considerable effort has been devoted to understanding both the mechanisms that produce polarized states and the conditions under which polarization can be reduced.

Many models of opinion dynamics focus on opinions regarding a single topic. However, individuals frequently hold correlated opinions on multiple issues, a phenomenon often referred to as opinion alignment \cite{ojer2023modeling,benoit2012dimensionality,baldassarri2008partisans,dimaggio1996have,schweighofer2020agent}. Motivated by this observation, Ojer et al. recently introduced the {\it social compass model} \cite{ojer2023modeling}, in which individuals hold opinions on two potentially correlated topics. The model exhibits a transition between polarized and consensus states as the strength of social interactions increases. Previous work focused primarily on the equilibrium properties of the model, on determining the critical coupling strength at which depolarization occurs, and on the conditions for an explosive depolarization transition.

The social compass model is structurally similar to the Kuramoto model of coupled phase oscillators \cite{kuramoto1975international}. In particular, the interaction term governing social influence in the social compass model has the same form as the coupling term in the Kuramoto model. This suggests that analytical techniques developed for oscillator synchronization, such as the Ott-Antonsen Ansatz \cite{ott2008low}, may also be useful for studying the dynamics of the social compass model. In this paper, we apply the Ott–Antonsen Ansatz to the social compass model and derive a reduced description of its macroscopic dynamics. This approach allows us to study not only the stationary state of the model but also the dynamical process of depolarization. We show that, for initially clustered opinion distributions, the dynamics reduce to a finite-dimensional system of ordinary differential equations. {From this reduced system of equations, we} derive a dispersion relation governing the growth of perturbations from polarized states and show that the critical coupling strength depends only on the first inverse moment of the conviction distribution, while the depolarization rate depends on higher moments. As a consequence, conviction distributions with identical depolarization thresholds can exhibit dramatically different relaxation times. Finally, we demonstrate how the framework can be extended to populations with community structure and use it to investigate the effects of communities on depolarization dynamics.

 The organization of this paper is as follows. In Sec.~\ref{sec:background}, we introduce the social compass model proposed in Ref.~\cite{ojer2023modeling}. {In Sec.~\ref{sec:ansatz}, we apply the Ott-Antonsen Ansatz to the social compass model to derive a reduced description of the dynamics and study the linear stability of the polarized state when it is macroscopically balanced. In Sec.~\ref{sec:approximate}, we study the depolarization transition from initial states that are polarized but only approximately macroscopically balanced.} In Sec.~\ref{sec:generalized}, we introduce and study a generalization of the social compass model that includes community structure. Finally, in Sec.~\ref{sec:discussion}, we discuss our results. The code for this project is available at \url{https://github.com/CorbitSampson/DynamicsSocialCompassModel}.  

\section{\label{sec:background} The social compass model}

\subsection{Phase representation of correlated opinions}

In this paper, as in Ref.~\cite{ojer2023modeling}, we will focus on studying opinions on only two topics $X$ and $Y$ (opinion alignment of a larger number of topics has been studied, for example, in Ref.~\cite{ojer2025social}). We assume that the opinions on topics $X$ and $Y$ can be represented by real numbers $x$ and $y$ on the open interval $(-\infty,\infty)$, with larger positive (negative) values representing stronger agreement (disagreement) with the topic. In the social compass model, pairs of opinions are represented by the complex number $x + i y = \rho e^{i \theta}$. The {\it conviction} $\rho$ represents the overall strength of the opinion pair and the phase angle $\theta$ represents the {\it orientation} of the two opinions. This polar representation is convenient to model opinion alignment, i.e., correlations between the two opinions among a population.

To illustrate opinion alignment in the social compass model, consider the following examples. First, suppose that topic X is ``Abortion rights'' and topic Y is ``Gun control'', with each topic measured on $[-1,1]$, for simplicity, with $-1$ representing being against and $1$ representing being in favor. In practice, these topics tend to be correlated, and therefore one would expect two extreme opinion pairs: $(-1,-1)$ and $(1,1)$. Alternatively, consider an example where topic X is ``Gun control'' and topic Y is ``Pineapple on pizza'', again measured on $[-1,1]$ with the same convention. A priori, one would not expect to find individuals' stances on these topics to be correlated, which suggests that, in this case, there are four extreme opinion pairs: $(-1,-1)$, $(-1,1)$, $(1,-1)$, and $(1,1)$.

\begin{figure}[t]
    \centering
    \includegraphics[width=0.6\linewidth]{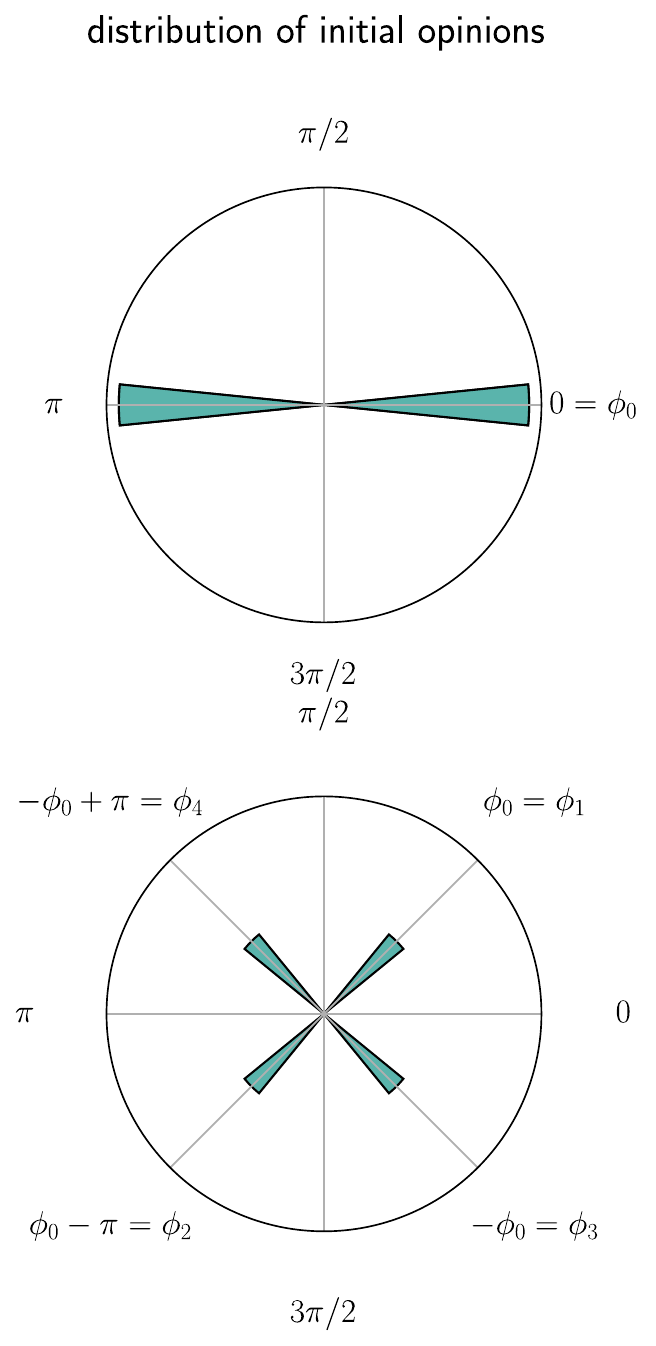}
    \caption[Polar histogram of initial opinion orientations]{A polar histogram of the distribution of initial opinion orientations for correlated initial opinions (top) and uncorrelated initial opinions (bottom).}
    \label{fig:quadrimodal}
\end{figure}

In each case, if one assumes that the population is split into equally sized factions of individuals that adopt all the possible extreme opinions,  then the distribution of opinion orientations for the first example is bimodal [see Fig.~\ref{fig:quadrimodal}(top), after a $-\pi/2$ rotation], while for the second example the distribution is quadrimodal [see Fig.~\ref{fig:quadrimodal}(bottom)].

These two examples represent {what is meant by ``polarized states" in the context of the social compass model}. The question addressed by the social compass model, formulated in the next section, is under what conditions initially polarized states evolve towards consensus or remain polarized.

\subsection{Model formulation}

In the social compass model~\cite{ojer2023modeling} it is assumed that $N$ agents hold opinions represented by their conviction and orientation, $\{(\rho_n, \theta_n)\}_{n=1}^N$, and that the opinion orientation of agent $n$ evolves as 
\begin{align}
    \frac{d\theta_n}{dt} &= \rho_n\sin(\phi_n - \theta_n) + \frac{K}{N}\sum _{m=1}^N\sin(\theta_m - \theta_n), \label{baseSCM}
\end{align}
where the angle $\phi_n = \theta_n(0)$ corresponds to the initial opinion orientation of agent $n$. The first term on the right-hand side of Eq.~(\ref{baseSCM}) represents a tendency of agent $n$ to revert to its own original opinion orientation, with a strength given by its conviction $\rho_n$. The second term represents a tendency to seek consensus with the other agents, which is modeled with a Kuramoto-type interaction term with coupling strength $K$.
We will assume that the initial agent orientations $\phi_n$ are randomly and independently drawn from a distribution $Q(\phi)$, while their convictions $\rho_n$ are similarly drawn from a distribution $P(\rho)$.  

Since consensus in this model is modeled as alignment of the orientations $\theta_n$, we can measure consensus as the magnitude of the Kuramoto complex-valued order parameter
\begin{align}
    z = re^{i\psi} &= \frac{1}{N} \sum _{n=1}^N e^{i \theta_n}.\label{SCPorderparameter}
\end{align}
We refer to the magnitude $r$ of this complex order parameter as the degree of consensus.  

Ref.~\cite{ojer2023modeling} shows that the social compass model exhibits a phase transition from a state of initially polarized opinions to one of consensus that is strongly dependent on the initial distribution of opinion orientations. For clustered initial conditions, the critical coupling and the order of the transition depend on the angular separation $\phi_0$ between opinion clusters. As  $\phi_0$ varies, the system can exhibit either continuous or discontinuous transitions to consensus. Additionally, Ref.~\cite{ojer2025social} shows that when coupling among individuals occurs along a network structure, then the critical coupling strength, marking the transition between polarization and consensus, depends strongly on the network topology. 

The method used in Ref.~\cite{ojer2023modeling} is to consider the thermodynamic limit $N\rightarrow \infty$ and solve for the equilibrium value of the order parameter using a self-consistent approach similar to the one used in early analyses of the  Kuramoto model \cite{acebron2005kuramoto}. However, as for the original Kuramoto model, the method of self-consistent equations does not give information about dynamical properties such as the stability of solutions or the timescale of depolarization. In the remainder of this paper, we will employ the Ott-Antonsen Ansatz introduced in Ref.~\cite{ott2008low} to study the social compass model. 

\section{\label{sec:ansatz} Dimensionality reduction of the social compass model}

Eq.~\eqref{baseSCM} can be rewritten in terms of the order parameter in Eq.~\eqref{SCPorderparameter} and the complex number representation of the initial state of each agent, $\rho_ne^{i \phi _n}$, as 
\begin{align}
    \frac{d\theta_n}{dt}&= \text{Im}\{(\rho_ne^{i \phi _n}+Kz)e^{-i\theta_n} \}.    \label{complexformSCP}
\end{align}
The form of the social compass model shown in Eq.~\eqref{complexformSCP} conforms to the case where the  Ott-Antonsen Ansatz \cite{ott2008low} can be applied. We begin by expressing the system's order parameter (\ref{SCPorderparameter}) in integral form by taking the continuum limit,
\begin{align}
    z &= \int _0^\infty \int _0^{2\pi} \int _0^{2\pi}f(t,\theta,\rho,\phi)e^{i\theta}d\theta d\phi d\rho, \label{contOPfull}
\end{align}
where $f(t,\theta,\rho,\phi)$ is the density of agents with opinion orientation $\theta$, conviction $\rho$, and initial opinion orientation $\phi$ at time $t$.
Since no agents are created or destroyed, $f$ satisfies the continuity equation
\begin{align}
    \frac{\partial f}{\partial t} + \frac{\partial}{\partial \theta}\left[\text{Im}\{(\rho e^{i\phi} + K  z)e^{-i\theta_i} \}f \right]&= 0.\label{continuityEQ}
\end{align}
We expand the density $f$ in Fourier series in $\theta$
\begin{align}
        f(t,\theta,\rho,\phi)&=\frac{P(\rho)Q(\phi)}{2\pi}\left[1 + \sum_{k=1}^\infty \alpha_k(t,\rho,\phi)e^{ik\theta} + \text{c.c.} \right],\label{densityseries}
\end{align}
with expansion coefficients $\alpha_k(t,\rho,\phi)$ and with $\text{c.c.}$ representing the complex conjugate of the preceding terms. We then make the Ansatz \cite{ott2008low} $\alpha_k(t,\rho,\phi) = [\alpha(t,\rho,\phi)]^k$ and insert Eq.~\eqref{densityseries} into the continuity equation [Eq.~\eqref{continuityEQ}]. Collecting like terms results in the same ODE for every $k$, 
\begin{align}
    \frac{d\alpha}{dt}(t,\rho,\phi)+\frac{\rho e^{i\phi} + K  z}{2}\alpha^2(t,\rho,\phi)-\frac{\rho e^{-i\phi} + K \bar z}{2}&=0,\label{reducedSCPgeneral}
\end{align}
where the bar indicates complex conjugation. Finally, inserting Eq.~\eqref{densityseries} into Eq.~\eqref{contOPfull} reduces the order parameter to
\begin{align}
    z&=\int _0^\infty \int_0^{2\pi}\overline{\alpha}(t,\rho,\phi)P(\rho)Q(\phi)d\phi d\rho.\label{reducedOPgeneral}
\end{align}
The resulting integro-differential equation described by Eqs.~\eqref{reducedSCPgeneral} and \eqref{reducedOPgeneral} fully describes the dynamics of the original system in the limit $N \to \infty$. 

To find the equilibrium solution, we start by proposing a uniformly rotating solution, $\alpha(t,\rho,\phi)= \alpha(\rho,\phi) e^{i \Omega t}$, where $\Omega$ is constant. Inserting this in Eqs.~(\ref{reducedSCPgeneral}) and (\ref{reducedOPgeneral}), one finds that such a solution is not possible unless $\Omega = 0$. This is expected since the conviction term breaks the rotational symmetry. Therefore, we can set $\partial\alpha/\partial t=0$ in Eq.~\eqref{reducedSCPgeneral}. Doing so, we find that the equilibrium solution is given by
\begin{align}
    \tilde \alpha  = \frac{\rho e^{-i\phi} + K\bar z}{|\rho e^{-i\phi} + K \bar z|}.\label{alphaequil}
\end{align}
Inserting this in Eq.~\eqref{reducedOPgeneral}, we find a self-consistent equation for the order parameter in steady state given by
\begin{align}
    z&=\int _0^\infty \int _0^{2\pi}P(\rho)Q(\phi) \frac{\rho e^{i\phi} + K z}{|\rho e^{i\phi} + K  z|}d\phi d\rho.\label{zequilgeneral}
\end{align}
This self-consistent equation is equivalent to Eq.~(3) in Ref.~\cite{ojer2023modeling}. 

\subsection{Finite-dimensional reduction for discrete opinion distributions}

To illustrate how the dynamics of the social compass model can be reduced in some cases to a finite set of ordinary differential equations, we consider a distribution of opinion orientations of the form 
\begin{align}\label{discrete}
Q(\phi) &= \sum_{k=1}^M d_k \delta(\phi - \phi_k),
\end{align}
where $d_k > 0$ and $\sum_{k=1}^M d_k = 1$. This distribution represents a situation where the population is initially divided into $M$ opinion clusters, with a fraction $d_k$ of the population having opinion orientation $\phi_k$.
This form includes the quadrimodal distribution 
\begin{align}
Q(\phi) &= \frac{1}{4}[\delta(\phi - \phi_0) + \delta(\phi + \phi_0) \label{quadrimodal}\\
& +\delta(\phi - \phi_0 +\pi) +\delta(\phi + \phi_0 - \pi)]\nonumber
\end{align}
that was studied in Ref.~\cite{ojer2023modeling}. For the conviction distribution, we assume for simplicity that $P(\rho) = \delta(\rho -1)$. With these choices, Eqs.~\eqref{reducedSCPgeneral} and \eqref{reducedOPgeneral} become [after evaluating Eq.~(\ref{reducedSCPgeneral}) at $\phi = \phi_k$] the finite system of ordinary differential equations
\begin{align}
    \frac{d\alpha_k}{dt} + \frac{e^{i \phi_k} + Kz}{2}\alpha_k^2 - \frac{e^{-i \phi_k} + K\;\overline{z}}{2}&=0, \quad k=1,2,\dots, M,\label{reducedSCPquadM}
\end{align}
and
\begin{align}
    z&=\sum_{k=1}^M d_k\overline{\alpha_k},\label{reducedOPquadM}
\end{align}
where $\alpha_k(t) = \alpha(t,1,\phi_k)$.
In particular, for the quadrimodal distribution (\ref{quadrimodal}), the system reduces exactly to a system of four coupled complex ordinary differential equations{, which is addressed explicitly in appendix \ref{appendix:B}.}

\subsection{\label{sec:stability} Linear stability of the macroscopically balanced state}

In this section, we use Eqs.~(\ref{reducedSCPgeneral})-(\ref{reducedOPgeneral}) to study the stability of a class of opinion states that we refer to as {\it macroscopically balanced states}. These states are defined by a distribution of opinion orientations $Q(\phi)$ such that the order parameter vanishes at time $t = 0$. In the continuum limit,
\begin{align}
  z(0) =   \int_0^{\infty} \int _0^{2\pi} e^{i\phi}Q(\phi)P(\rho) d\phi d\rho&= 0,\label{symmprop0}
\end{align}
which reduces to 
\begin{align}
\int _0^{2\pi} e^{i\phi}Q(\phi)d\phi&= 0.\label{symmprop}
\end{align}
Both the uncorrelated and correlated orientation distributions shown in Fig.~\ref{fig:quadrimodal} are macroscopically balanced. However, macroscopically balanced states do not necessarily correspond to a population divided into a finite number of polarized groups: the distribution $Q(\phi) = 1/(2\pi)$ is also macroscopically balanced. Although macroscopically balanced states require the exact condition (\ref{symmprop}), we will relax this assumption later on and study distributions that are {nearly} macroscopically balanced.

For simplicity, we will often use the term {\it polarized state} to refer to a macroscopically balanced state. While not all macroscopically balanced states are polarized, we are primarily interested in the transition from states like those shown in Fig.~\ref{fig:quadrimodal}, which satisfy $z = 0$, to a state of consensus, for which $|z| \sim 1$. However, we emphasize that the following results apply to any initial distribution of opinion orientations such that $z(0) = 0$.

\begin{figure*}[t]
    \centering
\includegraphics[width=1\linewidth]{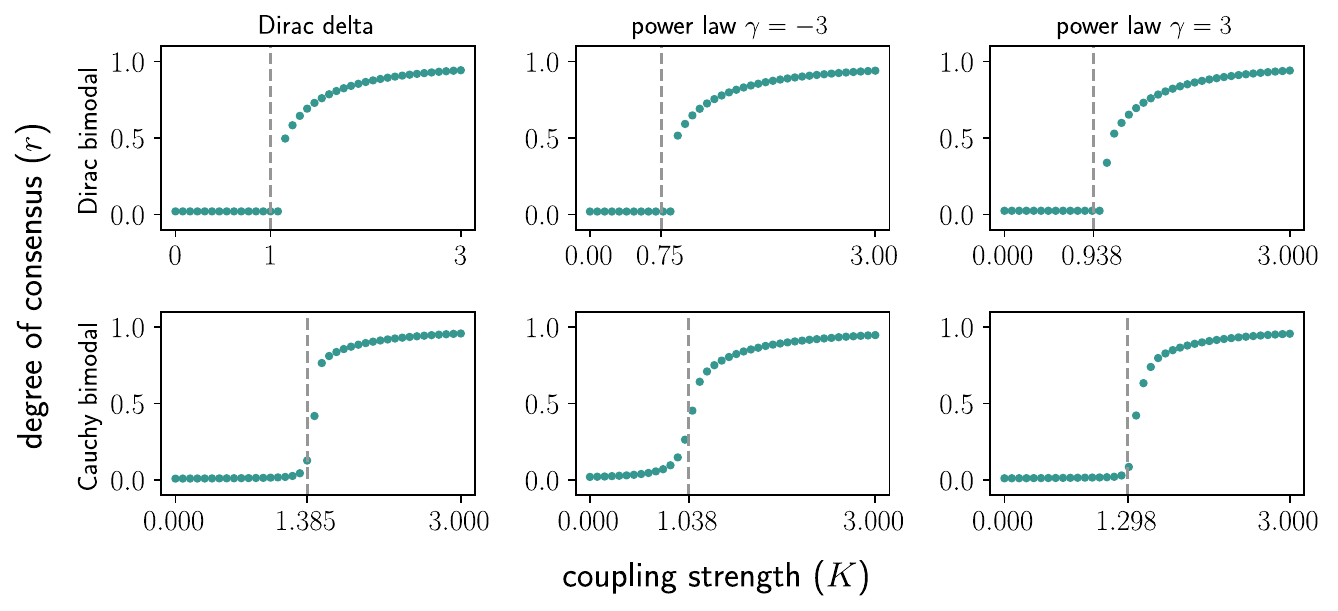}
    \caption[Forward continuation of the degree of consensus $r$ vs. the coupling strength $K$ for various initial distribution of opinion orientation and conviction]{Forward continuation of the degree of consensus $r$ plotted against the coupling strength $K$ for an initial distribution of opinion orientations $Q(\phi)$ given by Eq.~\eqref{quadrimodal} with {$\phi_0=0$} {(top row)}, which reduces to a bimodal Dirac delta distribution{,} and {a} {bimodal Cauchy} distributions [Eq.~\eqref{Cauchy}] (bottom row). The distributions of individuals' conviction $P(\rho)$ are Dirac delta distributions [Eq.~\eqref{diracdelta}] (first column), inverse power-law distribution [i.e., $P(\rho) \sim \rho^{-\gamma}$] (second column), and power-law distribution [i.e., $P(\rho)\sim \rho^\gamma$] (third column).}
    \label{critical_strength}
\end{figure*}

For macroscopically balanced states, $z = 0$ is a solution of the self-consistent equation (\ref{zequilgeneral}): setting $z = 0$ on both sides we get, after simplification,
\begin{align*}
    0 =\int _0^{\infty} P(\rho) d\rho \int _0^{2\pi} e^{i\phi}Q(\phi)d\phi = 0,
\end{align*}
where the final equality comes from Eq.~(\ref{symmprop}). This solution is characterized by [see Eq.~(\ref{alphaequil})]
\begin{align}
\alpha = e^{-i\phi}.
\end{align}
We study the linear stability of this solution by analyzing the time evolution of small perturbations 
\begin{align}
    \delta \alpha &= \alpha - e^{-i\phi}.
\end{align}
Linearizing Eq.~\eqref{reducedSCPgeneral} about $\alpha = e^{-i\phi}$ we get
\begin{align}
   \frac{1}{\rho} \frac{\partial \delta \alpha}{\partial t} + \delta \alpha + \frac{K}{2}\left[\frac{e^{-2i\phi}}{\rho}\delta z - \frac{1}{\rho}\overline{\delta z} \right]&=0,\label{reducedSCPlinear}
\end{align}
where [see Eq.~\eqref{reducedOPgeneral}]
\begin{align}
    \delta z&=\int _0^{\infty} P(\rho)\int _0^{2\pi} Q(\phi)\overline{\delta \alpha}\,d\phi d\rho. \label{eq:deltaz}
\end{align}
Anticipating exponential growth or decay of the perturbations, we set $\delta \alpha = a e^{\lambda t}$, where $\lambda$ and $a$ are to be found from the resulting eigenvalue problem. From Eq.~(\ref{eq:deltaz}), we have $\delta z = \bar a e^{\lambda t}$. Inserting these expressions in Eq.~(\ref{reducedSCPlinear}) and simplifying, we get
\begin{align}
   a (\lambda+\rho) + \frac{K}{2}\left[e^{-2i\phi}\bar a - a\right]&=0,\label{predispersion}
\end{align}
Dividing by $\rho + \lambda$, multiplying by $P(\rho)Q(\phi)$, and integrating over $\rho$ and $\phi$, we obtain
\begin{align}
  a = \frac{K}{2}\left(a - \bar a E_{\phi}[e^{-2i \phi}] \right) \int_{0}^{\infty}\frac{P(\rho)d\rho}{\rho + \lambda},\label{predispersion2}
\end{align}
where $E_{\phi}[\cdot]$ indicates an average over the distribution $Q(\phi)$.
This dispersion relation still depends on the eigenvector $a$. Separating  $E_{\phi}[e^{-2i \phi}]$ and $a$ into their real and imaginary parts, $E_{\phi}[e^{-2i \phi}] = u + i v$, $a = x + i y$, inserting these into Eq.~(\ref{predispersion2}), and separating the real and imaginary parts,  we obtain 
\begin{align}
\left(
\begin{array}{c}
x\\
y
\end{array}
\right) = \frac{K}{2}\int_{0}^{\infty}\frac{P(\rho)d\rho}{\rho + \lambda}
\left(
\begin{array}{cc}
1-u & -v\\
-v & 1+u
\end{array}
\right)
\left(
\begin{array}{c}
x\\
y
\end{array}
\right),
\end{align}
which implies 
\begin{align}
\frac{K}{2}\int_{0}^{\infty}\frac{P(\rho)d\rho}{\rho + \lambda} = \frac{1}{\Lambda},
\end{align}
where $\Lambda$ is an eigenvalue of the {above matrix}.
The largest eigenvalue, corresponding to the fastest growing mode, is $\Lambda = 1 + \sqrt{u^2+v^2} = 1 + |E_{\phi}[e^{-2i \phi}]|$. Therefore, we obtain the dispersion relation for the fastest growing mode,
\begin{align}
  1 = \frac{K}{2}\left(1 +  |E_{\phi}[e^{-2i \phi}]| \right) \int_{0}^{\infty}\frac{P(\rho)d\rho}{\rho + \lambda}.\label{dispersion1}
\end{align}
The eigenvector associated with the fastest growing mode, giving the direction in which the population will globally orient, is proportional to $(x,y)^T = (-v ,u + \sqrt{u^2+v^2})^T$. The corresponding complex amplitude $a = x + i y$, therefore, can be written as
\begin{align}
a = i\left(|E_{\phi}[e^{-2i \phi}]| + E_{\phi}[e^{-2i \phi}]\right). \label{eq:direction}
\end{align}
[In the degenerate case when $E_{\phi}[e^{-2i \phi}]$ is real and negative, one should use the eigenvector $(1,0)^T$, in which case $a = 1$.]
Since $a$ determines the perturbation $\delta z$, its complex argument specifies the direction in the complex plane along which the order parameter destabilizes. Importantly, this result gives the direction in which consensus emerges. Given an initial distribution of opinion orientations $Q(\phi)$ such that $z = 0$, if the system depolarizes and consensus emerges, it will do so by aligning the global opinion orientation in the direction given by Eq.~(\ref{eq:direction}). As an example, consider the quadrimodal distribution from Ref.~\cite{ojer2023modeling}. For this distribution, $E_{\phi}[e^{-2i \phi}]$ is real, and therefore the system orients in the imaginary direction when $E_{\phi}[e^{-2i \phi}]>0$, and in the real direction when $E_{\phi}[e^{-2i \phi}] < 0$.

Letting $\lambda \to 0^+$ in Eq.~(\ref{dispersion1}), we find the critical coupling strength for the onset of depolarization
\begin{align}
    K_c&= \frac{2 }{E_\rho[\rho^{-1}](1 + |E_\phi[e^{-2i\phi}]|)},\label{continuousKC}
\end{align}
which is in agreement with Eq.~(4) in Ref.~\cite{ojer2023modeling}. Equation~(\ref{continuousKC}) implies that the distribution of convictions only impacts the onset of depolarization via the expected value of $\rho^{-1}$. 

To illustrate this result, Fig.~\ref{critical_strength} shows the degree of consensus $r$ plotted against the coupling strength $K$ with a {dashed} vertical line marking the critical coupling strength $K_c$ [see Eq.~\eqref{continuousKC}] for six combinations of opinion orientation and conviction distributions. The top row corresponds to the bimodal case as in Eq.~\eqref{quadrimodal} with {$\phi_0=0$} and the bottom row to a bimodal Cauchy distribution wrapped onto the unit circle (see Appendix~\ref{bimodalcauchy}). For the distributions of individuals' convictions, we consider $\rho \equiv 1$ (left panels), an inverse power-law distribution with the limiting form $P(\rho) \sim \rho^{-3}$ above a minimum possible conviction $\rho_\text{min}$ (middle panels), and a truncated power-law distribution with the limiting form $P(\rho)\sim \rho ^3$ below a maximum possible conviction $\rho_\text{max}$ (right panels). We select the minimum conviction $\rho_\text{min}$ and the maximum conviction $\rho_\text{max}$ to ensure the same mean value $E_{\rho}[\rho] = 1$.

The degree of consensus $r$ is calculated using Eq.~\eqref{SCPorderparameter} from the results of a single numerical solution of Eqs.~\eqref{baseSCM} for each value of $K$ using $N=2000$ agents. Numerical solutions are produced using Euler's method with a step size $h=0.05$ for the time interval $[0,50]$. Initial conditions for $K=0$ are drawn for each agent randomly from the distribution $Q(\phi)$, and the initial conditions for subsequent values of $K$ are selected as the final state at $t=50$ from the previous value of $K$.

To demonstrate the additional insights provided by the Ott-Antonsen approach, we now study the relaxation time towards or away from the macroscopically balanced state near the critical point. For this, consider a coupling strength close to $K_c$, and let $K = K_c + \delta K$, where $\delta K \ll K_c$. Expanding the dispersion relation to first order in $\lambda$, we find to leading order that
\begin{align}
\lambda \approx \frac{\delta K}{K_c}\frac{E_\rho[\rho^{-1}]}{E_\rho[\rho^{-2}]}.
\end{align}
Therefore, two conviction distributions can lead to the same onset of depolarization $K_c$ (by virtue of sharing the same value of $E_\rho[\rho^{-1}]$), but have very different timescales of depolarization.

To illustrate this, we consider the family of conviction distributions given by 
\begin{align}
P_s(\rho) = \frac{s}{1+s} \delta(\rho - s) + \frac{1}{1+s}\delta\left(\rho -\frac{1}{s}\right),
\end{align}
where $s\geq 1$ parameterizes the heterogeneity in the conviction distribution while ensuring that $E_{\rho}[\rho^{-1}] = 1$. The distribution is completely homogeneous for $s = 1$, while it is composed mostly of stubborn agents with high conviction and a few agents with low conviction for large $s$. The variance of the conviction distribution is zero for $s = 1$, and diverges as $s \to \infty$.

\begin{figure}[t]
    \centering
\includegraphics[width=1\linewidth]{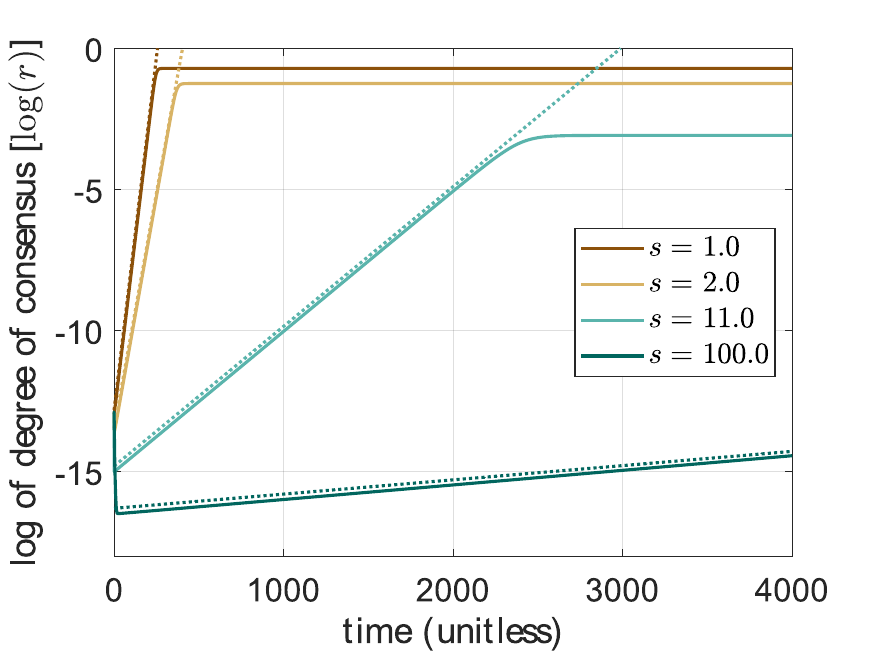}
    \caption{Logarithm of the order parameter $r$ versus time obtained from numerical simulations of Eqs.~(\ref{baseSCM}) with $s = 1, 2, 11$, and $100$, using a quadrimodal distribution of {initial opinion} orientations with $\phi_0 = \pi/7$ as solid lines. The dashed lines are drawn with the slope predicted by Eq.~(\ref{slope}) and an arbitrary intercept.}
    \label{rates}
\end{figure}

For this distribution, $E_{\rho}[\rho^{-2}] = s^2-s + 1/s$, and so we find that the growth rate of the order parameter at a coupling strength $K = K_c(1+\Delta)$, with $\Delta \ll 1$, is given by
\begin{align}
\lambda \approx \frac{\Delta}{s^2-s + 1/s}. \label{slope}
\end{align}
(An exact expression valid for larger $\Delta$ can also be written down in this case.) The fastest growth rate is obtained in the homogeneous case, $s = 1$, while it approaches zero as $s \to \infty$. Figure \ref{rates} shows the logarithm of the degree of consensus $r$ versus time obtained from numerical simulations of Eqs.~(\ref{baseSCM}) with $s = 1, 2, 11$, and $100$, using a quadrimodal distribution of orientations with $\phi_0 = \pi/7$ as solid lines. The dashed lines are drawn with a slope given by Eq.~(\ref{slope}) and an arbitrary intercept. The simulations were done with $N = 12120$, $K = 1.05K_c$, Euler's method with time step $\Delta t = 0.02$, and initial conditions where $\theta_n(0) = \phi_n + \epsilon_n$, where $\phi_n$ is sampled from the distribution $Q(\phi)$ and $\epsilon_n$ is a perturbation, chosen to be a zero-mean Gaussian with standard deviation $10^{-4}$. 

The theory predicts the growth rates very well, as evidenced by the slope of the {$\log[r(t)]$} plots coinciding with the slope of the dashed lines. As pointed out above, the rate of depolarization (i.e., the growth rate of the order parameter) changes by orders of magnitude as $s$ changes from $s = 1$ to $s = 100$, even though all these values of $s$ lead to eventual depolarization beyond $K = K_c$.

\section{Approximate macroscopically balanced states}\label{sec:approximate}

In Sec.~\ref{sec:stability} we focused our attention on the linear stability of the no-consensus state $z = 0$, which is possible only for distributions satisfying  (\ref{symmprop}). In practice, while one would not expect this condition to be satisfied exactly, one might still expect a depolarization transition from states with $z \approx 0$. Here we will tackle this problem by assuming that the distribution of {opinion} orientations $Q(\phi)$ is approximately macroscopically balanced. Specifically, we will consider a distribution of initial opinion orientations $Q(\phi)$ such that 
\begin{align}
    Q(\phi) &= \overline{Q}(\phi) + q(\phi),
\end{align}
where 
\begin{align}
    \int _0^{2\pi}Q(\phi)d\phi =  \int _0^{2\pi}\bar Q(\phi)d\phi = 1,\label{eq:norms}
\end{align}
$\overline{Q}(\phi)$ satisfies the condition
\begin{align}
    \int _0^{2\pi}\overline{Q}(\phi)e^{i\phi}d\phi = 0, \label{eq:mbs3}
\end{align}
and $q(\phi)$ represents a small perturbation in the distribution of initial opinion orientations (note that $q$ must integrate to zero). The introduction of $q(\phi)$ will result in a distribution of initial opinion orientations that is ``unbalanced'', resulting in a nonzero value of $r$ at equilibrium. To find this value, we perform an expansion of the order parameter in Eq.~\eqref{zequilgeneral} about the equilibrium $\tilde \alpha$ in Eq.~\eqref{alphaequil}. For simplicity, we again assume that $P(\rho)=\delta(\rho - 1)$. 

We begin with the order parameter equation [see Eq.~\eqref{zequilgeneral}] at equilibrium, i.e., 
\begin{align*}
    z&= \int _0^{2\pi}\frac{e^{i\phi} + Kz}{|e^{i\phi} + Kz|}Q(\phi)d\phi,
\end{align*}
or
\begin{align}
    z&= \int _0^{2\pi}\frac{[e^{i\phi} + Kz][\overline{Q}(\phi)+q(\phi)]}{(1 + K(e^{i\phi}\overline{z} + e^{-i\phi}z) + K^2|z|^2)^{1/2}}d\phi.
\end{align}
Now, noting that for a small perturbation $q(\phi)$ the order parameter $z$ will also be small, we can make an expansion of the denominator about $z = 0$ to get
\begin{align}
    z&\approx \int _0^{2\pi} [e^{i\phi} + Kz]\left[1 - \frac{K}{2}(e^{i\phi}\overline{z}+e^{-i\phi}z)\right]\left[\overline{Q}(\phi)+q(\phi)\right]d\phi.
\end{align}
Next, distributing the integrand, dropping higher-order terms, and applying Eqs.~\eqref{eq:norms} and ~\eqref{eq:mbs3}, we get
\begin{align}
    z&\approx - \frac{K}{2} E_{\bar Q}[ e^{2i \phi}]\overline{z} +\frac{K}{2}z+ \langle e^{i\phi},q\rangle,
\end{align}
where $\langle f,g\rangle = \int_0^{2\pi} f(\phi) g(\phi)d\phi$. By moving to a rotated coordinate system $\theta \to \theta + \beta$, $\phi \to \phi + \beta$, we can{,} without loss of generality{,} assume that $E_{\bar Q}[ e^{2i \phi}]$ is real and nonnegative, since this transformation results in $E_{\bar Q}[ e^{2i \phi}] \to e^{2i\beta} E_{\bar Q}[ e^{2i \phi}]$. Letting $b = E_{\bar Q}[ e^{2i \phi}]$ and $c e^{i \gamma} = \langle e^{i\phi},q\rangle$, and $z=re^{i\psi}$, we have
\begin{align}
    re^{i\psi}&= -\frac{Kbr}{2}e^{-i\psi}+\frac{Kr}{2}e^{i\psi}+ce^{i\gamma}.
\end{align}
Splitting this equation into its real and imaginary parts results in the pair of equations
\begin{align}
    r\cos(\psi)&=\frac{c\cos(\gamma)}{[1+\frac{Kb}{2}-\frac{K}{2}]},\label{cospsi}\\
    r\sin(\psi)&=\frac{c\sin(\gamma)}{[1-\frac{Kb}{2}-\frac{K}{2}]}\label{sinpsi}.
\end{align}
Squaring Eqs.~\eqref{cospsi} and \eqref{sinpsi} and adding them together results in the following expression for the degree of consensus
\begin{align}
    r&=c \sqrt{\frac{\cos^2(\gamma)}{[1+\frac{K}{2}(1-b)]^2} + \frac{\sin^2(\gamma)}{[1-\frac{K}{2}(1+b)]^2}}.\label{rfrompert}
\end{align}

Eq.~\eqref{rfrompert} shows that when the distribution of initial opinion orientations deviates from a macroscopically balanced state by a small perturbation $q$, the magnitude of the order parameter scales linearly with the size of the perturbation, i.e., with $c = |\langle e^{i\phi},q\rangle|$. Furthermore, Eq.~\eqref{rfrompert} contains two singularities, due to the two denominators. The first occurs at 
\begin{align}
K_c=\frac{2}{1+b}, \label{singularityKC}
\end{align}
which agrees with Eq.~(\ref{continuousKC}),
and the second at $K_2=2/(1-b)$ (recall that $b = |E_{\bar Q}[ e^{2i \phi}]| \geq 0$). Since $K_c\leq K_2$, $r$ will  diverge as $K\rightarrow K_c$ if $\sin(\gamma) \neq 0$ [if $\sin(\gamma) = 0$, the singularity at $K = K_c$ is absent and the divergence of $r$ occurs at $K = K_2$ instead]. Since Eq.~\eqref{rfrompert} was derived under the assumption that $r$ is small, we interpret $K_c$ as the coupling strength resulting in a transition from a small non-consensus equilibrium to a state of consensus. 

To validate this result, we now consider an example where
\begin{align}
    Q(\phi) &= (1-c) \bar Q(\phi) + c\delta (\phi - \gamma),\label{pert_distribution}
\end{align}
where $\bar Q$ is the quadrimodal distribution in Eq.~(\ref{quadrimodal}) with $\phi_0 = \pi/6$. In Eq.~\eqref{pert_distribution}, $c$ represents the size of the perturbation and $\gamma$ the angle in the complex plane at which it occurs, and therefore these variables correspond to the $c$ and $\gamma$ used throughout the derivation of Eq.~(\ref{rfrompert}). We consider the three values $c = 0$, $0.04$, and $0.2$, {$\gamma=\pi/4$,} and $\rho \equiv 1$, and compare the predictions of Eq.~(\ref{rfrompert}) with the agent-based model (\ref{baseSCM})
and the steady-state value of the reduced system (\ref{reducedSCPgeneral})-(\ref{reducedOPgeneral}). The other parameters used are as follows. For the agent-based model, we consider a collection of $N=1000$ agents and solve Eq.~\eqref{baseSCM} via Euler's method with step size $\Delta t=0.05$ for $t\in (0,200]$ with a phase offset $\phi_0=\pi/6$ with $K\in [0,2.5]$. {We use the same parameters and conventions for the reduced system [Eqs.~\eqref{reducedSCPgeneral}-\eqref{reducedOPgeneral}].} 

\begin{figure}[t]
    \centering
    \includegraphics[width=1\linewidth]{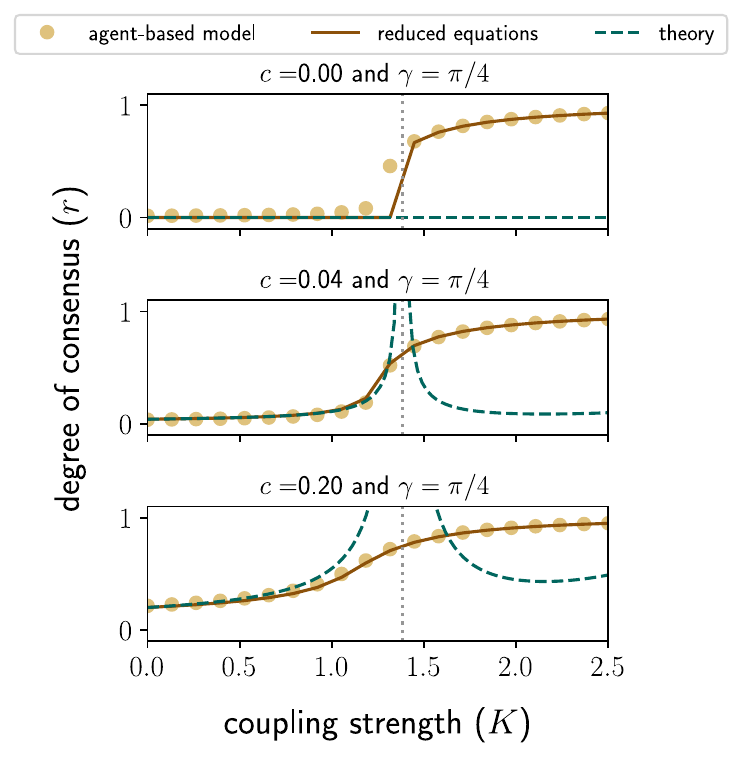}
    \caption{Degree of consensus $r$ versus the coupling strength $K$ with {$\gamma=\pi/4$ and $\rho=1$ for} $c = 0$ (top), $c = 0.04$ (middle), and $c = 0.2$ (bottom). The circles show the value of $r$ obtained from Eq.~(\ref{baseSCM}), the solid lines the steady state of Eqs.~(\ref{reducedSCPgeneral})-(\ref{reducedOPgeneral}), the dashed lines the value of $r$ from Eq.~(\ref{rfrompert}){, and the vertical dotted lines mark $K_c$ from Eq.~\eqref{singularityKC}.}}
    \label{pert_continuation}
\end{figure}

In Fig.~\ref{pert_continuation} we plot the degree of consensus $r$ versus the coupling strength $K$ for $c = 0$ (top), $c = 0.04$ (middle), and $c = 0.2$ (bottom). The circles show the value of $r$ obtained from Eq.~(\ref{baseSCM}), the solid lines the steady-state of Eqs.~(\ref{reducedSCPgeneral})-(\ref{reducedOPgeneral}), and the dashed lines the value of $r$ from Eq.~(\ref{rfrompert}). When $c=0$, the distribution is macroscopically balanced and there is a stable solution with $r = 0$ for low values of $K$. Introducing a small perturbation in the distribution of initial opinion orientations ($c > 0$) creates a stable steady state where the degree of consensus is nonzero. For small enough values of $c$ (e.g., $c=0.04$ in the middle panel), the value of $r$ for low values of $K$ is close to zero, and it transitions to larger values at values of $K$ close to $K_c$. The transition becomes less well defined for larger values of $c$, but $K_c$ still provides a good approximation to the location of the transition region.

\section{\label{sec:generalized} Community structure}

In this section we extend the social compass model to include the effects of community structure, illustrating the flexibility of the Ott-Antonsen Ansatz to handle generalizations of the basic social compass model. 

Community structure often plays an important role in social interactions, as it is often a network structure that partially (or fully) isolates individuals from different groups. Community structure is usually defined in terms of the number of connections between individuals within a community relative to the number of connections out of the community \cite{newman2006modularity}. Here, we will consider a somewhat similar approach where the network is all-to-all, but intracommunity coupling is stronger than intercommunity coupling.

\subsection{Derivation of reduced equations with communities} 

We consider a generalization of the social compass model where there are two communities, labeled $1$ and $2$, {each} of sizes $N/2$, where $N$ is the total number of agents. The model is described by the following equation:
\begin{align}
    &\frac{d\theta_n^\sigma}{dt} = \rho_n^\sigma \sin(\phi_n^\sigma - \theta_n^\sigma) + \frac{K^{\sigma \sigma}}{N}\sum _{m=1}^{N/2}\sin(\theta_m^\sigma - \theta_n^\sigma)\notag\\
    &+ \frac{K^{\sigma \sigma'}}{N}\sum _{m=1}^{N/2}\sin(\theta_m^{\sigma'} - \theta_n^\sigma), \hspace{0.5cm} n = 1,2,\dots, N/2,\label{SCM_CF} 
\end{align}
where $\theta_n^\sigma$ and $\rho_n^\sigma$ are the opinion orientation and conviction, respectively, of node $n$ in community $\sigma$. The phase angle $\phi_n^\sigma = \theta_n^\sigma(0)$ is the initial opinion orientation of node $n$, $K^{\sigma \sigma}$ is the intracommunity coupling, and $K^{\sigma \sigma'}$ is the intercommunity coupling (assuming $\sigma \neq \sigma'$). 

As in the case without community structure, we can define an order parameter to measure the amount of consensus in each community. We define [see Eq.~\eqref{SCPorderparameter}]
\begin{align}
    z_\sigma = \frac{2}{N}\sum_{m=1}^{N/2}e^{i \theta_m^\sigma},\label{localOP}
\end{align}
as the local order parameter of community $\sigma$, and a global order parameter given by
\begin{align}
    z&=\frac{1}{2}\left(z_\sigma + z_{\sigma'} \right).\label{localtoglobalOP}
\end{align}
With the definition of the local order parameter in Eq.~\eqref{localOP}, we can express the generalized social compass model in Eq.~\eqref{SCM_CF} as
\begin{align}
    \frac{d\theta_n^\sigma}{dt}&=\text{Im}\left\{H_\sigma e^{-i \theta_n^\sigma}\right\}, \hspace{0.5cm} n = 1,2,\dots, N/2,
\end{align}
where $\sigma \in \{1,2\}$ and
\begin{align}
    H_\sigma &= \rho_n^\sigma e^{i\phi_n^\sigma} + \frac{1}{2}K^{\sigma \sigma} z_\sigma + \frac{1}{2}K^{\sigma \sigma'}z_{\sigma '}.\label{Hcomm} 
\end{align}

Applying the Ott-Antonsen Ansatz, as in Sec.~\ref{sec:ansatz}, we arrive at the reduced description of the system dynamics given by
\begin{align}
    \frac{d\alpha_\sigma}{dt}&= \frac{1}{2}\overline{H}_\sigma - \frac{1}{2}H_\sigma \alpha_\sigma ^2,\label{reducedcommunity}
\end{align}
for each community $\sigma$ and 
\begin{align}
    z_\sigma &= \int _0^\infty \int _0^{2\pi} \overline{\alpha}_\sigma(t,\rho,\phi)P_\sigma(\rho)Q_\sigma(\phi) d\phi d\rho,\label{continuouscommunityOP}
\end{align}
where $P_\sigma(\rho)$ is the distribution of convictions and $Q_\sigma(\phi)$ is the distribution of initial opinion orientations of individuals in community $\sigma$. Lastly, we can write down a self-consistent equation for the equilibrium solutions of the generalized social compass model. The equilibrium solution to Eq.~\eqref{reducedcommunity} is
\begin{align}
    \alpha_\sigma &= \frac{\overline{H_\sigma}}{|\overline{H_\sigma}|}.
\end{align}
Substituting the equilibrium solution into Eq.~\eqref{continuouscommunityOP} we get
\begin{align}
    z_\sigma &= \int _0^\infty \int _0^{2\pi} \frac{{H_\sigma}}{|{H_\sigma}|}(t,\rho,\phi)P_\sigma(\rho)Q_\sigma(\phi) d\phi d\rho.\label{community_order_equil}
\end{align}
Finally, using Eq.~\eqref{localtoglobalOP} we have
\begin{widetext}
\begin{align}
    z&=\frac{1}{2}\int_0^\infty\int _0^{2\pi} \left[\frac{{H_1}}{|{H_1}|}P_1(\rho)Q_1(\phi) + \frac{{H_{2}}}{|{H_{2}}|}P_{2}(\rho)Q_{2}(\phi) \right]d\phi d\rho.\label{global_order_equil}
\end{align}
\end{widetext}
Together, Eqs.~\eqref{global_order_equil} and \eqref{Hcomm} form a self-consistent system of equations for the global order parameter at equilibrium. 

\subsection{A quadrimodal distribution and critical coupling strength}

\begin{figure}[t]
    \centering
    \includegraphics[width=0.7\linewidth]{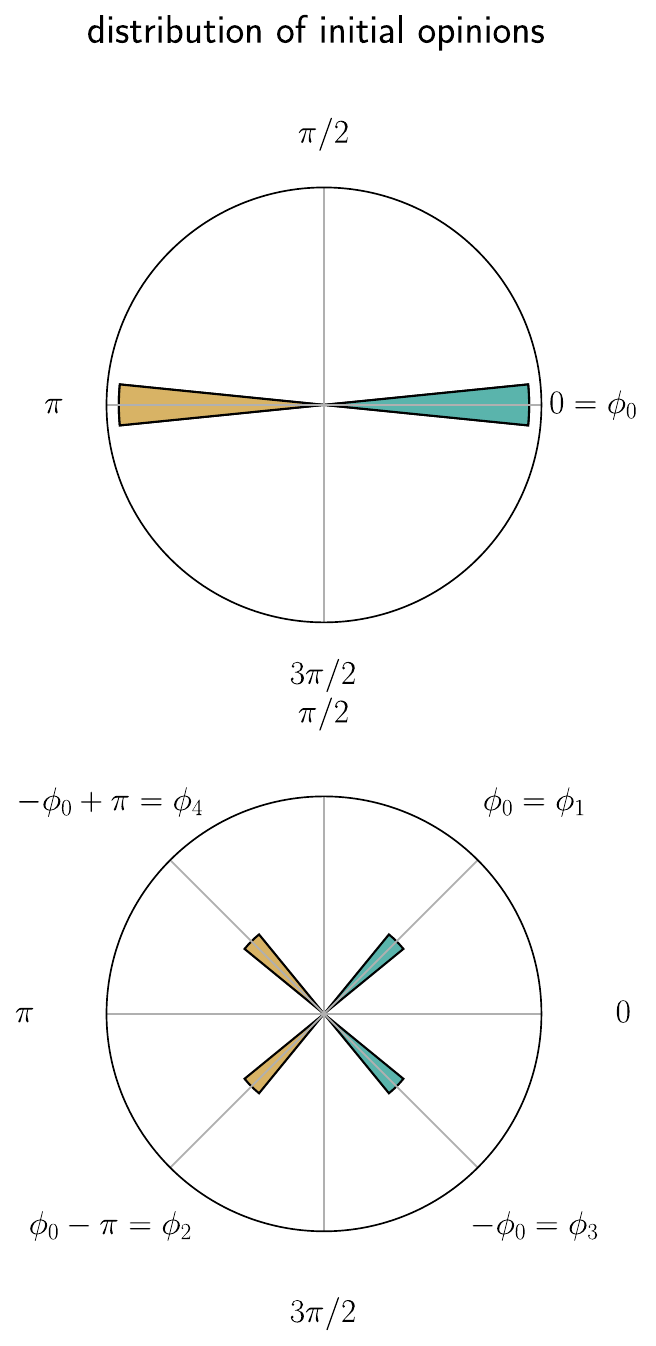}
    \caption[Polar histogram of initial opinion orientations with communities]{A polar histogram of the distribution of initial opinion orientations for correlated initial opinions (top) and uncorrelated initial opinions (bottom). The blue bars correspond to initial opinion orientations for community $1$ and the tan bars correspond to initial opinion orientations for community $2$.}
    \label{fig:quadrimodal_comm}
\end{figure}

For simplicity, we will consider a variation of the two-community case where the distribution of initial opinion orientations is quadrimodal. In particular, we take the distributions of initial opinion orientations for communities $1$ and $2$, respectively, to be
\begin{align}
    Q_1(\phi) &= \frac{1}{2}\left[\delta(\phi - \phi_0) + \delta(\phi + \phi_0) \right], \label{comm1}
\end{align}
and
\begin{align}
     Q_{2}(\phi) &= \frac{1}{2}\left[\delta(\phi - \phi_0 + \pi) + \delta(\phi + \phi_0 - \pi) \right]. \label{comm2}
\end{align}
In Fig.~\ref{fig:quadrimodal_comm}, we include a schematic representation of these distributions with the modes associated with community $1$ in blue and the modes associated with community $2$ in tan.
This choice is made to ensure that initially correlated opinions ($\phi_0 = 0$) are held in distinct communities. We also make the choice that the intercommunity coupling ($K^{\sigma \sigma'}$ with $\sigma'\neq \sigma$) has the form $K^{\sigma \sigma'}=\mu K$ with $0\leq \mu \leq 1$, and that the intracommunity coupling is $K^{\sigma \sigma}=K$. The choice $\mu=0$ corresponds to completely disconnected communities, while $\mu = 1$ corresponds to no community structure. 

Using the self-consistent equations for the local and global order parameters given by Eqs.~\eqref{community_order_equil} and \eqref{global_order_equil}, respectively, we can now perform an expansion to determine the critical coupling strength that defines the transition to consensus.

Because the distribution of initial opinion orientations in community $2$ is obtained from that of community $1$ by the transformation $\phi \to \pi - \phi$, the equations are invariant under the corresponding transformation $z_1 \to -\bar{z}_2$. We therefore restrict attention to the symmetry-preserving branch satisfying
\begin{align}
    z_{\sigma}&= -\bar{z}_{\sigma '}. \label{symmetry}
\end{align}
We now consider a state in which each community reaches a {partial internal consensus} ($z_1 \neq 0$, $z_2 \neq 0$), but there is no global {consensus (i.e., $z = 0$)}. Eq.~(\ref{symmetry}), together with $z = (z_1 + z_2)/2 = 0$, implies that both of the order parameters are real,
\begin{align}
    z_1 &= r,\\
    z_2 &= -r.
\end{align}
Now we consider a perturbation from this solution:
\begin{align}
    z_1 &= r + \delta z_1, \label{sig_expand1}\\
    z_2 &= -r +\delta z_2 .\label{sig_expand2}
\end{align}
Substituting the initial opinion orientations, given in Eqs.~\eqref{comm1} and \eqref{comm2}, into Eq.~\eqref{community_order_equil}, we have 
\begin{align}
    z_{1}&= \frac{1}{2}\left[\frac{e^{i\phi_0} + \frac{1}{2}K(z_1 + \mu z_{2})}{|e^{i\phi_0} + \frac{1}{2}K(z_1 + \mu z_{2})|} + \frac{e^{-i\phi_0} + \frac{1}{2}K(z_1 + \mu z_{2})}{|e^{-i\phi_0} + \frac{1}{2}K(z_1 + \mu z_{2})|} \right],\nonumber\\ 
    z_{2}&= \frac{1}{2}\left[\frac{e^{-i\phi_0} + \frac{1}{2}K(z_2 + \mu z_{1})}{|e^{-i\phi_0} + \frac{1}{2}K(z_2 + \mu z_{1})|} + \frac{e^{i\phi_0} + \frac{1}{2}K(z_2 + \mu z_{1})}{|e^{i\phi_0} + \frac{1}{2}K(z_2 + \mu z_{1})|} \right]
    \label{two_community_local_order}.
\end{align}
First, to zeroth order in $\delta z_1$, $\delta z_2$, inserting $z_1 = r$, $z_2 = -r$ in Eq.~(\ref{two_community_local_order}), we get
\begin{align}
    r&=\frac{R + \cos(\phi_0)}{[1+R^2+2R\cos(\phi_0)]^{1/2}}\label{equilibriumRsigma},
\end{align}
where
\begin{align}
R=\frac{K(1-\mu)r}{2}.\label{bigR}
\end{align}
Substituting Eq.~(\ref{bigR}) into Eq.~(\ref{equilibriumRsigma}), one obtains a self-consistent equation for $r$, which can be solved numerically given $K$, $\mu$, and $\phi_0$. Now we consider an expansion to first order in $\delta z_1$, $\delta z_2$. Linearizing Eq.~(\ref{two_community_local_order}), one finds
\begin{align}
\delta z_1 &= A(\delta z_1 + \mu \delta z_2) +  B(\overline{\delta z_1} + \mu \overline{\delta z_2})\label{linearsystem1},\\
\delta z_2 &= A(\delta z_2 + \mu \delta z_1) +  B(\overline{\delta z_2} + \mu \overline{\delta z_1})\label{linearsystem2},
\end{align}
where 
\begin{align}
A &= \frac{K}{4[1+R^2+2R\cos(\phi_0)]^{1/2}},\\
B &= -\frac{K \left[ R^2 + 2 R \cos(\phi_0) + \cos(2\phi_0)\right]}{4 [1+R^2+2R\cos(\phi_0)]^{3/2}}.
\end{align}
Equations (\ref{linearsystem1})-(\ref{linearsystem2}), together with their complex conjugates, can be viewed as a linear system for the variables $(\delta z_1, \delta z_2,\overline{\delta z_1},\overline{\delta z_2})$. Setting the largest eigenvalue of the matrix associated with the right-hand side of this system to one gives the critical value of the coupling strength as
\begin{align}
    K_{c}&= \frac{2 [1+R^2+2R\cos(\phi_0)]^{3/2}}{(1+\mu)[1+R^2+2R\cos(\phi_0) - \sin^2(\phi_0)]}.\label{criticalKComm}
\end{align}
To determine the critical coupling strength, we use the value of $R$ obtained from solving Eqs.~\eqref{equilibriumRsigma} and \eqref{bigR}.

\begin{figure}
    \centering
    \begin{subfigure}{0.47\textwidth}
        \includegraphics[width=\linewidth]{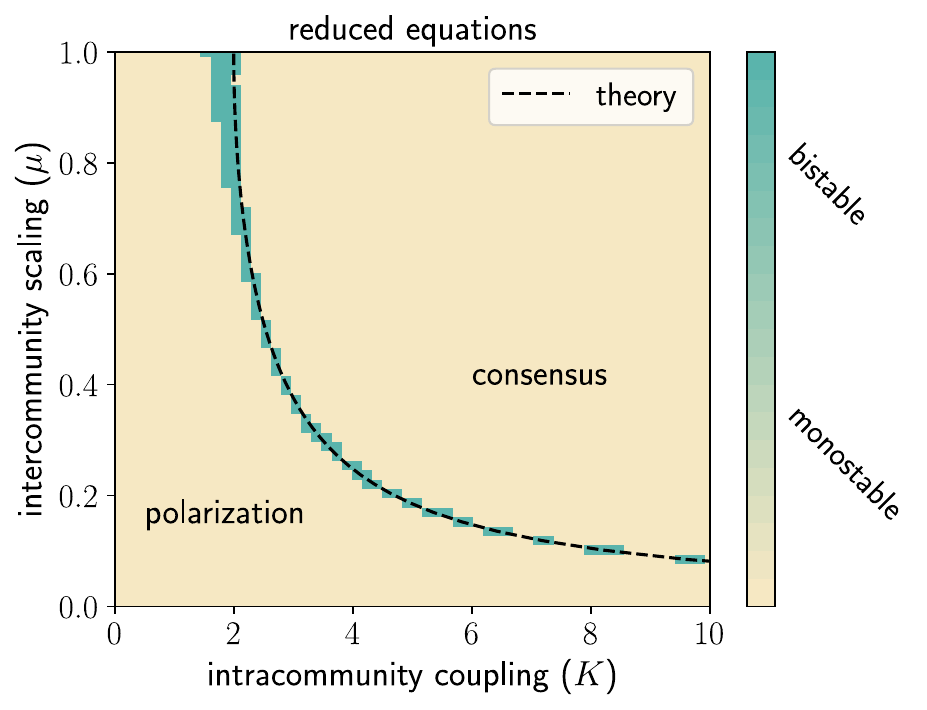}
    \end{subfigure}
    \hfill
    \begin{subfigure}{0.47\textwidth}
        \includegraphics[width=\linewidth]{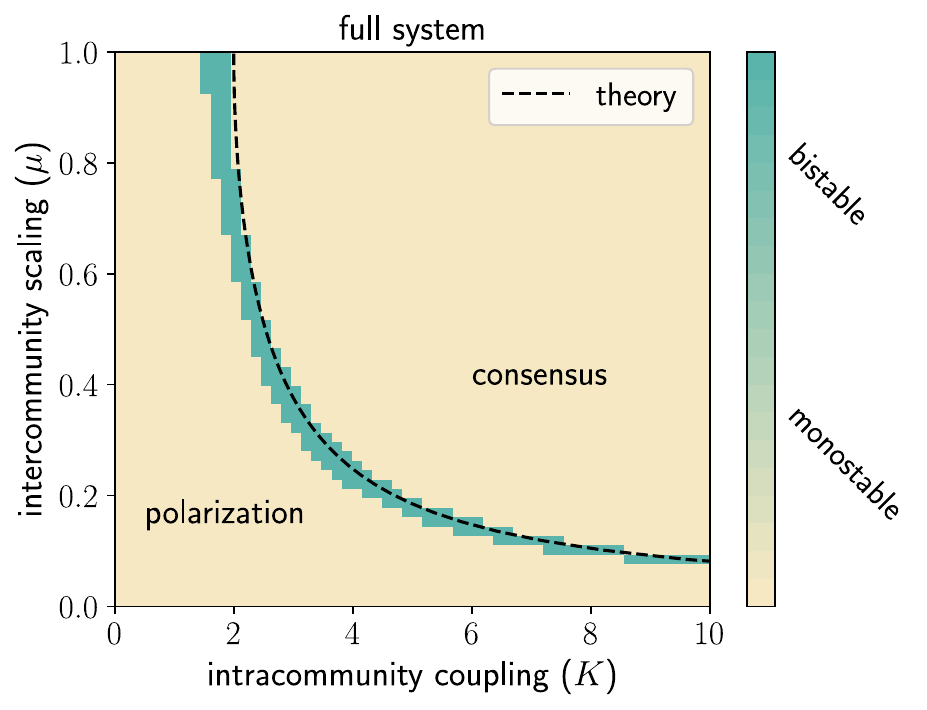}
    \end{subfigure}
    \caption[Bifurcation diagrams for the social compass model with communities]{\label{bistable_mu_Kin} (Top) A bifurcation diagram of the steady-state behavior of Eq.~\eqref{reducedcommunity} in the region $\mu\in [0,1]$ and $K\in[0,10]$ for two communities with initial opinion distributions as in Eqs.~\eqref{comm1} and \eqref{comm2} with $\phi_0 = \pi/4$ and convictions all equal to $1$ [i.e., as in Eq.~\eqref{diracdelta} with $\rho_\text{max} = 1$].  (Bottom) A bifurcation diagram of the steady-state behavior of the agent-based model [i.e., Eq.~\eqref{SCM_CF}] with $N=1500$ in the region $\mu\in [0,1]$ and $K\in[0,10]$ for the same conventions as in the reduced equations [i.e., Eq.~\eqref{reducedcommunity}].}
\end{figure}

We now compare the predictions for the onset of global depolarization obtained using the reduced equations [Eq.~\eqref{reducedcommunity}], the original social compass model [Eq.~\eqref{SCM_CF}], and our prediction of the critical coupling strength [Eq.~\eqref{criticalKComm}]. Fig.~\ref{bistable_mu_Kin} shows a bifurcation diagram for the transitions between a polarized state ($r=0$), a consensus state ($r\approx 1$), and a bistable state versus $K$ and $\mu$ for {the reduced description of the dynamics [top, Eq.~\eqref{reducedcommunity}] and the agent-based model [bottom, Eq.~\eqref{SCM_CF}]}; the black dashed curve indicates the critical coupling strength $K_c$ computed via a numerical solution to the system of equations comprised of Eqs.~\eqref{equilibriumRsigma} and \eqref{criticalKComm}. We see that as the amount of intercommunity coupling decreases, reaching consensus requires larger intracommunity coupling, and the region of bistability becomes narrower. Additionally, comparing the top and bottom panels of Fig.~\ref{bistable_mu_Kin}, we see good qualitative agreement between the reduced-dimensional description of the dynamics [Eq.~\eqref{reducedcommunity}] and the agent-based model [Eq.~\eqref{SCM_CF}]. We also see that the theoretical critical coupling strength describes well the boundary between bistability and monostable consensus.

In Fig.~\ref{bistable_mu_Kin} (top and bottom), we observe that the boundary between consensus and polarization seems to approach an asymptote $\mu_{\infty}$ as $K\rightarrow \infty$. The value of this asymptote determines whether sufficiently strong {intracommunity} coupling allows even very weak intercommunity coupling to align the communities. If $\mu_{\infty} > 0$, then the polarized state will not destabilize into consensus for weak enough intercommunity coupling, $\mu < \mu_{\infty}$. On the other hand, if $\mu_{\infty} = 0$, strong enough intracommunity coupling will lead to destabilization of the polarized state even for very weak intercommunity coupling.

To determine the value of $\mu_{\infty},$ we consider the limiting behavior of Eq.~\eqref{criticalKComm} as $K\to \infty$ and $\mu \to \mu_{\infty}$ along the curve $(K_c(\mu),\mu)$. {Note that, from Eq.~\eqref{bigR}, $R\propto K$}. {Combining this with} Eq.~\eqref{equilibriumRsigma} {we find} that 
\begin{align}
    \lim _{K\rightarrow \infty} r &= \lim _{K\rightarrow \infty} \frac{R + \cos(\phi _0)}{[1+R^2+2R\cos(\phi_0)]^{1/2}}=1.\notag
\end{align}
Next, we consider the behavior of $K_c$ as $K\to \infty$ along the $(K_c(\mu),\mu)$ curve. Using Eq.~(\ref{criticalKComm}) and the definition of $R$ in Eq.~(\ref{bigR}), we find
\begin{align}
  \lim _{K\rightarrow \infty} K_c &=  \lim _{K\rightarrow \infty}\frac{2 [1+R^2+2R\cos(\phi_0)]^{3/2}}{(1+\mu)[1+R^2+2R\cos(\phi_0) - \sin^2(\phi_0)]} \\
  &= \lim _{K\rightarrow \infty}\frac{2R}{1+\mu}\\
    &=\frac{(\lim _{K\rightarrow \infty}K_c)(1-\mu_{\infty})}{1+\mu_{\infty}}.
\end{align}
Balancing the equations, we find $1=(1-\mu_{\infty})/(1+\mu_{\infty})$.
Solving for $\mu_{\infty}$, we find $\mu_{\infty} = 0$. Therefore, we see that there is an asymptote at $\mu=0$ that is approached in the limit as $K_c \rightarrow \infty$. This means that for any positive value of $\mu$ there will always exist a finite $K_c$ that acts as a critical coupling strength, resulting in the transition to consensus for $K>K_c$. 

\section{\label{sec:discussion} Discussion}

In this paper, we have shown that the social compass model admits a low-dimensional description by using the Ott-Antonsen Ansatz in the limit $N \to \infty$. This approach is complementary to the self-consistent equilibrium analysis of Ref.~\cite{ojer2023modeling} and allows the dynamics of the model to be studied directly. In particular, we derived reduced equations governing the evolution of the order parameter and showed that, for discrete distributions of opinion orientations, the infinite-dimensional dynamics reduce exactly to a finite system of coupled ordinary differential equations.

Using this reduced description, we also studied the stability of a broad class of macroscopically balanced states characterized by zero initial consensus. Linearization about these states yielded a dispersion relation that determines both the critical coupling strength associated with the onset of consensus and the relaxation or growth rate of perturbations. This extends previous analyses of the social compass model by providing explicit information about the dynamics, rather than only the equilibrium states.

One consequence of this analysis is that the onset of consensus and the rate at which consensus forms are controlled by different properties of the conviction distribution. While the critical coupling strength depends only on the mean inverse conviction $E_{\rho}[\rho^{-1}]$, the growth rate depends on higher inverse moments. Therefore, populations with identical consensus thresholds can approach consensus on dramatically different timescales. Within the family of conviction distributions considered here, populations with homogeneous conviction distributions approach consensus at the highest rate. This result highlights the importance of conviction heterogeneity in determining not only whether consensus is possible, but also how rapidly it emerges.

We also examined the effects of small asymmetries in the distribution of initial opinion orientations. We found that such asymmetries give rise to equilibrium states with nonzero consensus and smooth out the transition between polarized and consensus states. Although the location of the critical coupling strength remains unchanged to leading order, consensus can begin to emerge before the critical point. This suggests that even small minority groups that break an otherwise exact polarization can significantly alter the collective dynamics.

Finally, we generalized the social compass model to include community structure and derived a corresponding low-dimensional description. Our analysis shows that weaker intercommunity coupling suppresses consensus formation and reduces the size of the bistable region. In the limit of vanishing intercommunity coupling, global consensus becomes impossible, while stronger intercommunity coupling promotes alignment across communities. 

The framework developed here opens several directions for future work. Possible extensions include the study of more complex community structures, networked interactions, time-dependent conviction distributions, and adaptive social networks. More generally, the Ott-Antonsen approach provides a systematic route for analyzing the dynamics of social compass models beyond equilibrium and may prove useful in understanding other opinion-formation processes with heterogeneous individual tendencies.

\bibliography{SCM_V02}

\appendix

\section{Bimodal Cauchy distribution}\label{bimodalcauchy}

\begin{figure}[t]
    \centering
    \includegraphics[width=0.7\linewidth]{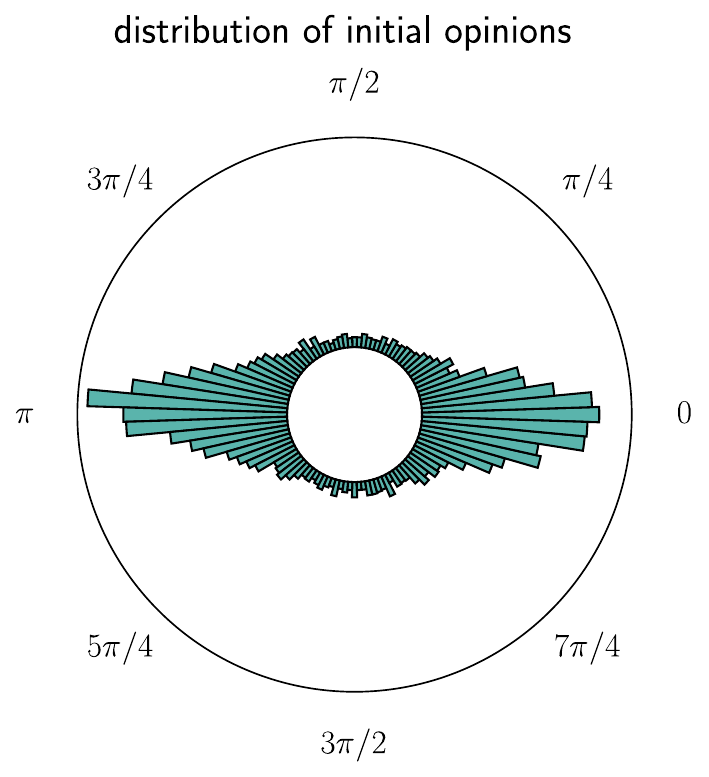}
    \caption[Example of Distribution in Eq.~\eqref{Cauchy}]{A polar histogram showing an example of the wrapped Cauchy distribution in Eq.~\eqref{Cauchy}.}
    \label{cauchy}
\end{figure}

The bimodal Cauchy distribution wrapped around the unit circle has a  probability density function of the form
\begin{widetext}
\begin{align}
    Q(\phi)&=\frac{1-v^2}{4\pi} \left(\frac{1}{1+v^2-2v\cos(\phi - \mu_1)} + \frac{1}{1+v^2-2v\cos(\phi-\mu_2)}\right),\label{Cauchy}
\end{align}
\end{widetext}
{where the $\mu_i$'s represent the center of each mode and $v$ is a scale factor for the distribution.} Figure \ref{cauchy} shows an example of a distribution of {angles $\phi$ sampled from such $Q(\phi)$ using $v = 1/1.3$ with $\mu_1=0$ and $\mu_2 = \pi$.}

\section{\label{appendix:B}Recovery of the results of Ref.~\cite{ojer2023modeling}}

In this section, we consider one of the two cases considered in Ref.~\cite{ojer2023modeling}. In particular, we examine the case where the distribution of individuals' convictions is
\begin{align}
    P(\rho) = \delta(\rho - \rho_{\text{max}}),\label{diracdelta}
\end{align}
and the distribution of initial opinion orientations is 
\begin{align}
    Q(\phi) &= \frac{1}{4}\left [\delta(\phi - \phi_{0}) + \delta(\phi - \phi_{0} + \pi)\right.\nonumber \\
    &\left.+ \delta(\phi + \phi_0) + \delta(\phi + \phi_0 - \pi) \right],\label{quad}
\end{align}
where $\phi_0 \in [0,\pi/4]$ parameterizes the transition from correlated initial opinions (at $\phi_0 = 0$) to uncorrelated initial opinions (at $\phi_0 = \pi/4$). The distribution $Q(\phi)$ is shown schematically in Fig.~\ref{fig:quadrimodal}. 

In this Appendix, using Eqs.~\eqref{reducedSCPgeneral} and \eqref{reducedOPgeneral}, we will reproduce the self-consistent equation of Ojer et al., from which all of their results can be recovered.  

To begin, we apply our choices of $P(\rho)$ and $Q(\phi)$ to Eqs.~\eqref{reducedSCPgeneral} and \eqref{reducedOPgeneral}, resulting in a collection of ordinary differential equations for the dynamics of $\alpha(t,\rho,\phi)$, coupled by the global order parameter $z$. With these choices, Eqs.~\eqref{reducedSCPgeneral} and \eqref{reducedOPgeneral} become
\begin{align}
    \dot{\alpha_n} + \frac{a_n + Kz}{2}\alpha_n^2 - \frac{\overline{a_n} + K\overline{z}}{2}&=0, \quad n=1,2,3,4,\label{reducedSCPquad}
\end{align}
and
\begin{align}
    z&=\frac{1}{4}\sum_{m=1}^4 \overline{\alpha_m},\label{reducedOPquad}
\end{align}
respectively, where $\alpha_n(t)=\alpha(t,\rho_{\text{max}},\phi_n)$, $a_n=\rho_{\text{max}}e^{i\phi_n}$, $\phi_1 = \phi_0$, $\phi_2 = \phi_0 - \pi$, $\phi_3 = -\phi_0$, and $\phi_4 = -\phi_0 +\pi$. Next we define $\alpha_n=r_ne^{i\eta_n}$ and split Eq.~\eqref{reducedSCPquad} into its real and imaginary parts. This results in 
\begin{align}
    \dot{r}_n + i r_n\dot{\eta}_n + \frac{(a_n + Kz)e^{i\eta_n}}{2}r_n^2 + \overline{\frac{(a_n + Kz)e^{i\eta_n}}{2}}&=0,
\end{align}
or
\begin{align}
    \dot{r}_n &= -\frac{1}{2}(r_n^2 - 1)\left[\rho_{\text{max}}\cos(\phi_n + \eta_n) + Kr\cos(\psi + \eta_n) \right],\label{radialSCPquad} \\ 
    r_n\dot{\eta}_n&=-\frac{1}{2}(r_n^2 + 1)\left[\rho_{\text{max}}\sin(\phi_n + \eta_n) + Kr\sin(\psi + \eta_n) \right],\label{angularSCPquad}
     \\ n&=1,2,3,4,\notag
\end{align}
where $z=re^{i\psi}$. From here, note that $\dot{r}_n=0$ whenever $r_n = 1$, an invariant space which is observed numerically to be stable. On the invariant surface where $r_n= 1$ for $n=1,2,3,4$, the dynamics of the phases reduce to
\begin{align}
    \dot{\eta}_n&=\rho_{\text{max}}\sin(-\phi_n - \eta_n) + Kr\sin(-\psi -\eta_n), \quad n=1,2,3,4.\label{angularSCPquadreal}
\end{align}
In the following, we will search for stationary solutions on this invariant surface.

\begin{figure}[t]
    \centering
    \includegraphics[width=1\linewidth]{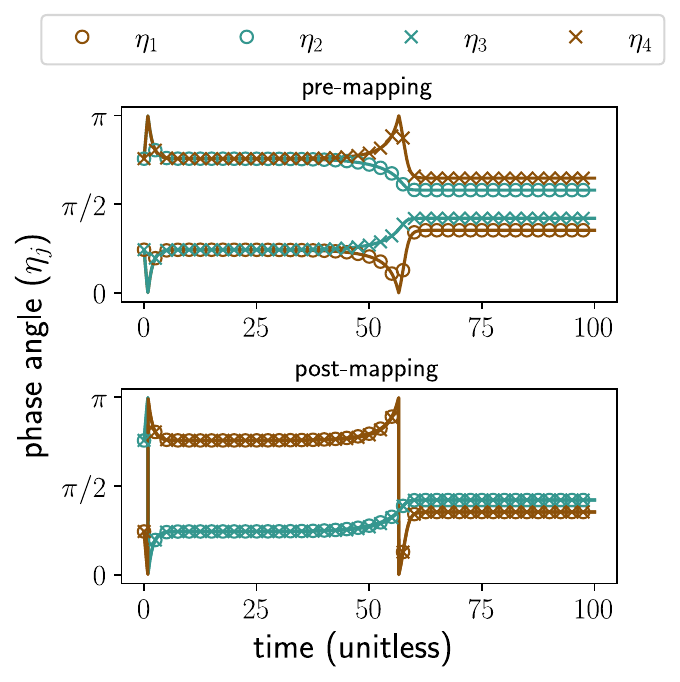}
    \caption[Remapping opinion orientations with Eqs.~\eqref{quadmapping1} and \eqref{quadmapping2}.]{The phases $\eta_j$ from numerical solutions of Eq.~\eqref{angularSCPquadreal} (top). The phases $\eta_j$ from numerical solutions of Eq.~\eqref{angularSCPquadreal} after applying the mappings in Eqs.~\eqref{quadmapping1} and \eqref{quadmapping2} (bottom). {For both cases the phases are restricted to the interval $[0,\pi]$ via the mapping $\eta_j \rightarrow \eta_j \,\text{mod} \pi$.}}
    \label{remapping}
\end{figure}

Next, we will make use of the symmetry about the imaginary axis in the distribution $Q(\phi)$ to further reduce the system in Eq.~\eqref{angularSCPquadreal}. Note that 
\begin{align}
    \eta_2 (0) &= \phi_2 = \pi - \eta_3(0),\label{quadmapping1}\\
    \eta_4 (0) &=\phi_4 =  \pi - \eta_1(0),\label{quadmapping2}
\end{align}
and that therefore, from Eq.~\eqref{reducedOPquad}, $\psi(0) = \pm \frac{\pi}{2}$. One can check that $\eta_2 + \eta_3$ and $\eta_1 + \eta_4$ are conserved by Eqs.~\eqref{angularSCPquadreal} as long as $\psi=\pm \frac{\pi}{2}$, and so we can assume that $\eta_2 + \eta_3 = -\pi$, $\eta_1 + \eta_4=\pi$, $\psi = \pm \frac{\pi}{2}$ and reduce the dimension of the system. Additionally, Fig.~\ref{remapping} shows the effect of applying the transformations $\eta_2 = -\pi - \eta_3$, $\eta_4 = \pi - \eta_1$ to a numerical solution of Eq.~\eqref{angularSCPquadreal}, further validating this claim. Making the substitutions $\eta_2 = -\pi - \eta_3$, $\eta_4 = \pi - \eta_1$ in Eq.~\eqref{reducedOPquad}, we obtain 
\begin{align}
    z &= \frac{1}{4}\left[e^{-i\eta_1} + e^{-i(\pi - \eta_3)} + e^{-i\eta_3} + e^{-i(\pi - \eta_1)}\right]\notag\\
    &=-\frac{i}{2}\left[\sin(\eta_1) + \sin(\eta_3) \right].\label{quadsymmetricZ}
\end{align}
Thus, the order parameter lies on the imaginary axis, meaning that $\psi = \pm \pi/2$. Enforcing this on Eq.~\eqref{angularSCPquadreal}, we find that
\begin{align}
    \dot{\eta}_n&=\rho_{\text{max}}\sin(-\phi_n - \eta_n) \pm Kr\cos(\eta_n), \quad n=1,2,3,4.
\end{align}
Lastly, we apply the transformation in Eqs.~\eqref{quadmapping1} and \eqref{quadmapping2} to get
\begin{align}
    \dot{\eta_1}&=\rho_{\text{max}}\sin(-\phi_0 - \eta_1) \pm Kr\cos(\eta_1),\label{quadsymmetricSCP1}\\
    \dot{\eta_3}&=\rho_{\text{max}}\sin(\phi_0 - \eta_3) \pm Kr\cos(\eta_3),\label{quadsymmetricSCP2}
\end{align}
with the order parameter $z$ defined by Eq.~\eqref{quadsymmetricZ}. Combining Eqs.~\eqref{quadsymmetricSCP1} and \eqref{quadsymmetricSCP2}, together with Eq.~\eqref{quadsymmetricZ}, we capture the dynamics of the original high-dimensional system restricted to the invariant surface $r_n=1$. 

Next, we want to use this result to recover the key results of Ojer et al. To do this, we solve Eqs.~\eqref{quadsymmetricSCP1} and \eqref{quadsymmetricSCP2} for the constant-valued solutions $\eta_1^*$, $\eta_3^*$, corresponding to $\dot{\eta}_1=0$ and $\dot{\eta}_3=0$. {(Note that we already excluded the possibility of rotating solutions in Sec.~\ref{sec:ansatz} because of the lack of rotational symmetry due to the conviction term.)} These are
\begin{align}
    \eta_1^*&=\arctan \left[-\tan(\phi_0) \pm \frac{Kr}{\rho_\text{max}}\sec(\phi_0)\right],\\
    \eta_3^*&=\arctan \left[\tan(\phi_0) \pm \frac{Kr}{\rho_\text{max}}\sec(\phi_0)\right].
\end{align}
Then, substituting these into $r=\frac{1}{2}\left[\sin(\eta_1^*) + \sin(\eta_3^*) \right]$ and simplifying, we get
\begin{align}
    r&= \frac{-\rho_\text{max} \sin(\phi_0) \pm Kr}{2\sqrt{\rho_\text{max}^2 + (Kr)^2 - Kr\rho_\text{max}\sin(\phi_0)}}\notag\\
    &+ \frac{\rho_\text{max} \sin(\phi_0) \pm Kr}{2\sqrt{\rho_\text{max}^2 + (Kr)^2 + Kr\rho_\text{max}\sin(\phi_0)}}.\label{quadselfconsistent}
\end{align}
The right-hand side of Eq.~\eqref{quadselfconsistent} is identical to equation (31) in the Supplemental Material of Ref.~\cite{ojer2023modeling}, meaning that additional results, such as the onset of bistability and the critical coupling strength, can be recovered using the same techniques from this point {forward}.

\end{document}